\documentclass[showpacs,aps,prl,twocolumn,superscriptaddress,10pt]{revtex4-2}
\usepackage{graphicx} 
\usepackage{bm}
\usepackage{color}

\usepackage{amsmath}
\usepackage{amssymb}
\usepackage{enumerate}
\usepackage{xspace}
\usepackage{mathrsfs}
\usepackage{mathptmx}
\usepackage[colorlinks=true,linkcolor=blue,citecolor=blue,urlcolor=blue]{hyperref}
\usepackage{orcidlink}
\usepackage[utf8]{inputenc}

\newcommand{\vampire}{\textsc{vampire} }

\newcommand{\kB}{\ensuremath{k_{\mathrm{B}}}\xspace}

\begin{document}

\title{Atomistic calculation of the $f_0$ attempt frequency in Fe$_3$O$_4$ magnetite nanoparticles}

\author{Roberto~Moreno\orcidlink{0000-0002-9799-4210}}
\affiliation{CONICET, Instituto de Física Enrique Gaviola (IFEG), Córdoba, Argentina}
\affiliation{School of GeoSciences, University of Edinburgh,  Edinburgh, UK}
\affiliation{School of Physics, Engineering and Technology, University of York, York, YO10 5DD, UK}
\author{Sarah~Jenkins\orcidlink{0000-0002-6469-9928}}
\affiliation{School of Physics, Engineering and Technology, University of York, York, YO10 5DD, UK}
\author{Wyn~Williams\orcidlink{0000-0001-9210-7574}}
\affiliation{School of GeoSciences, University of Edinburgh,  Edinburgh, UK}
\author{Richard~F.~L.~Evans\orcidlink{0000-0002-2378-8203}}
\affiliation{School of Physics, Engineering and Technology, University of York, York, YO10 5DD, UK}
\email{richard.evans@york.ac.uk}
\begin{abstract}

The Arrhenius law predicts the transition time between equilibrium states in physical systems due to thermal activation, with broad applications in material science, magnetic hyperthermia and paleomagnetism where it is used to estimate the transition time and thermal stability of assemblies of magnetic nanoparticles. Magnetite is a material of great importance in paleomagnetic studies and magnetic hyperthermia but existing estimates of the attempt frequency $f_0$ vary by several orders of magnitude in the range $10^7-10^{13}$ Hz, leading to significant uncertainty in their relaxation rate.
Here we present a dynamical method enabling full parameterization of the Arrhenius-N\'eel law using atomistic spin dynamics. We determine the temperature and volume dependence of the attempt frequency of magnetite nanoparticles with cubic anisotropy and find a value of $f_0 = 0.562 \pm 0.059$ GHz at room temperature. For particles with enhanced anisotropy we find a significant increase in the attempt frequency and a strong temperature dependence suggesting an important role of anisotropy. The method is applicable to a wide range of dynamical systems where different states can be clearly identified and enables robust estimates of domain state stabilities, with particular importance in the rapidly developing field of  micromagnetic analysis of paleomagnetic recordings where samples can be numerically reconstructed to  provide a better understanding of geomagnetic recording fidelity over geological time scales.
\end{abstract}

\maketitle

The Arrhenius law~\cite{Arrhenius1889}  predicts the mean time that an equilibrium state of a physical system is stable before thermal activation overcomes the energy barrier for transition into another equilibrium state. The law was originally developed in chemistry to explain and determine reaction frequencies in chemical processes~\cite{Arrhenius1889,arrhenius_svante_1889_1749766,doi:10.1021/ed061p494}, but due to its simplicity and versatility it is now used across a wide range of disciplines and research areas~\cite{doi:10.1021/ed049p343,ArrheniuosAPP,ArrheniuosAPP2}. 
One example is in nanomagnetism, where the Arrhenius-N\'eel law ~\cite{neel1949} can determine the average time scales for transitions between equilibrium magnetization states in magnetic nanocrystals. This is of special interest for a wide range of applications such as magnetic recording media \cite{MRM_ARR1}, medical sciences \cite{Papadopoulos2022-fm} or paleomagnetism \cite{neel1949}. In paleomagnetism, one important example is geomagnetic recording in the titanomagnetite series of minerals which are commonly found within rocks or meteorites, and often carry the bulk of the magnetic remanence. The present-day magnetic states in these naturally grown ancient nanocrystals are dependent on their thermo-chemical history as well as Earth's magnetic field intensity and direction at the time of their formation. These ancient recordings can tell us about key events during the evolution of the Earth and Solar System, such as early stagnant-lid tectonics~\cite{RN18525}, nucleation of the inner core~\cite{RN18528,RN18529} and the influence of the geomagnetic field on the atmospheric and creating conditions suitable for evolution of life~\cite{RN18526}, amongst others.

In contrast to the complexity of scientific problems to which it has been applied,  the form of the Arrhenius-N\'eel law is extremely simple: 
\begin{equation}
    \tau = \tau_0\exp\left({\frac{E_BV}{{k_BT}}}\right) = \frac{1}{f_0}  \exp\left({\frac{E_BV}{{k_BT}}}\right)
    \label{Arrenhius}
\end{equation}
where $\tau$ is the mean time a system stays in an local energy minima magnetic state (with energy $E_1$) before it rotates to a second state (with energy $E_2$) or vice versa.  $\tau_0$ is the characteristic switching time of a  magnetic nanostructure and its inverse $f_0 = 1/\tau_0$ the attempt frequency of the system. The relaxation time $\tau$ depends on the ratio of the energy barrier $E_\mathrm{B}V$ between the local energy minima states and the thermal energy $k_\mathrm{B}T$ of the system. $k_\mathrm{B}$ is the Boltzmann constant and $T$ is the absolute temperature of the system. In nanomagnetism, $E_\mathrm{B}$ is determined by the different types of anisotropies in the system, e.g. magnetocrystalline or shape anisotropy~\cite{Moreno2020}. In the particular case of isotropic magnetite nanoparticles cubic magnetocrystalline anisotropy determines the energy barrier as $E_\mathrm{B} = K_\mathrm{C}/12$, where $K_\mathrm{C}$ represents the negative cubic energy above the Verwey transition temperature. 

Different methods exist to calculate the energy barrier $E_\mathrm{B}$ of a magnetic nanostructure such as constrained Monte Carlo~\cite{PhysRevB.82.054415} or the Nudged elastic band method ~\cite{10.1093/gji/ggy285,RN18655,RN18657} applied beyond the single domain (SD) regime to larger vortex and multi-domain magnetite particles\cite{Nagy10356}. In contrast, determination of $\tau_0$ is more complicated and a wide range of values exist in the literature, spanning from $10^{-7} $ to $10^{-13}$ s for the case of iron oxide nanoparticle systems \cite{RN18513,PhysRevB.48.10240,DICKSON1993345,ATTEMP1,https://doi.org/10.1002/adfm.201101243}. The disparity in these values has important consequences for paleomagnetic recordings in the determination of whether a natural sample has kept its original magnetisation. In magnetic hyperthermia $\tau_0$ determines which nanoparticle size will optimize the specific absorption rate (SAR) \cite{https://doi.org/10.1002/adfm.201101243}. For simplicity, an approximate value for the attempt frequency of $1$ GHz has been widely assumed in the literature derived from the effects of magnetic relaxation in magnetic nanoparticles have on their M\"ossbauer sprectrum \cite{RN18531} but this required highly characterised experimental samples and the fit to theory requires several simplifications. Additionally, two independent analytical expressions for $f_0$ were reported in the literature by N\'eel \cite{neel1949} and Brown \cite{Brown} that are both volume and temperature dependent. In practice $f_0$ is considered a constant and whether the volume or temperature play an important role on the value of $f_0$ is still an open question.

\begin{figure}[!tb]
    \centering
    \includegraphics[width=0.95\columnwidth ,angle=0]{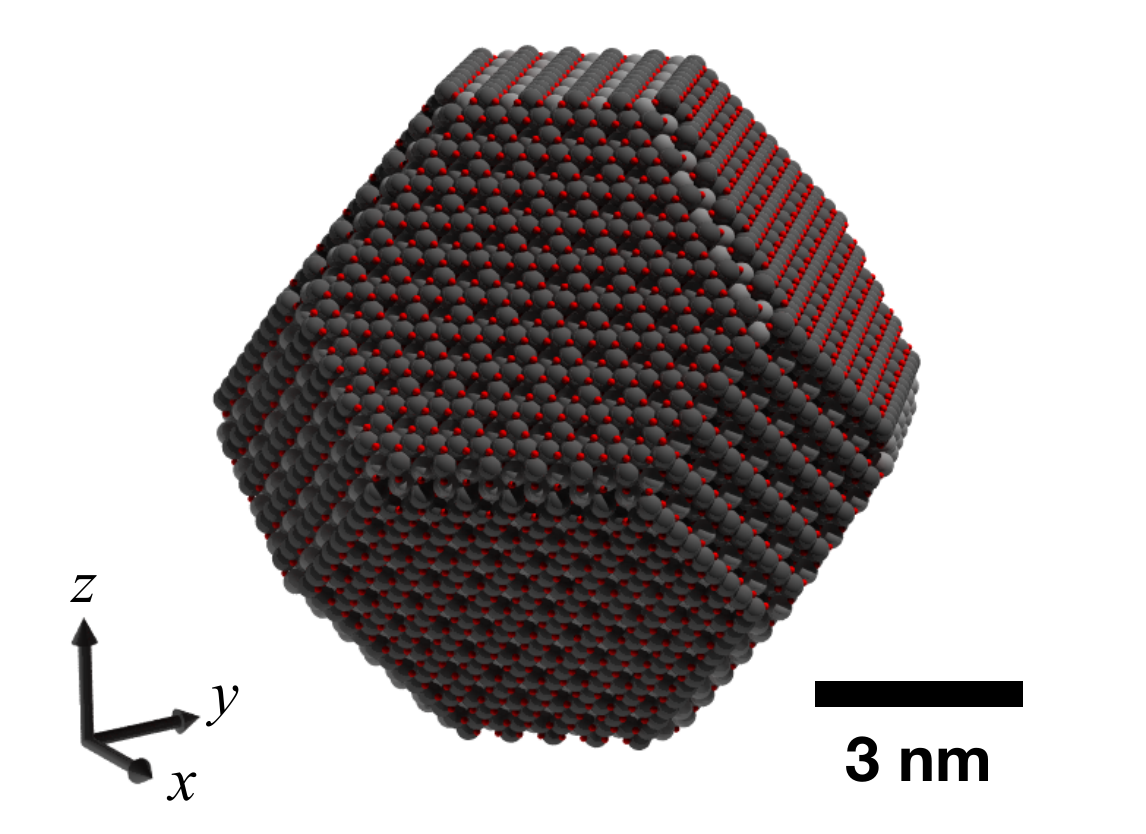}
    \caption{(a) Structure of an truncated octahedral faceted magnetite nanoparticle. Dark and light grey spheres represent the tetrahedral and octahedral Fe sublattices respectively, while red spheres represent the Oxygen atoms. The $x$,$y$ and $z$ directions show the $\langle$ 1 0 0 $\rangle$ directions with the  $\langle$111 $\rangle$ easy magneto crystalline directions orthogonal to the hexagonal planes. This particle corresponds a 9 nm diameter particle and has $\sim$ 15000 magnetic atoms.
    }
    \label{fig:schematic}
\end{figure}

In this letter we numerically determine an accurate value for $f_0$ of magnetite nanoparticles using atomistic spin dynamics which models the time evolution of the magnetization time in the presence of thermal fluctuations. The value of $f_0$ is calculated for magnetite nanoparticles of varying volumes and anisotropies and at different temperatures and we present an alternative strategy to determine the temperature dependence of the energy barrier $E_\mathrm{B}$ in magnetic systems.

To model the dynamics of magnetite nanoparticles we use a Heisenberg spin Hamiltonian 
\begin{equation}
\mathcal{H}=-\frac{1}{2} \sum_{i\neq j} J_{ij}
\mathbf{S}_i \cdot \mathbf{S}_j + \frac{1}{2}\sum_i k_{\mathrm{c}} \left(S_{i,x}^4 + S_{i,y}^4  + S_{i,z}^4 \right), 
\label{Heis}
\end{equation}
which considers the magnetic moments to be localized at the magnetic atoms \cite{vampire}. $\mathbf{S}_i$ is a unit vector representing the direction of the magnetic moment of the atom at site $i$. $J_{ij}$ is the exchange energy between the magnetic moments located at $i$ and $j$ respectively where exchange interactions are mediated by Fe-O-Fe super-exchange bonds. $k_{\mathrm{c}}$ represents the cubic magnetocrystalline anisotropy energy experienced by the magnetic moment $i$. Exchange parametrization is taken from \cite{Moreno_2021} where 
it was successfully used to describe the main bulk properties and anomalous magnetic phenomena in magnetite thin films with the presence of antiphase boundaries (APBs).
We consider two distinct cases of magnetic anisotropy considering fully passivated Fe$_3$O$_4$ surfaces with bulk magnetic anisotropy $k_{\mathrm{c}} = -5.033 \times 10^{-25}$ J/atom which reproduces the correct magnetic anisotropy energy at room temperature of $K_\mathrm{C}(300K) = -1.1 \times 10^{4}$ J/m$^3$ \cite{PhysRevB.46.5334} when considering temperature rescaling~\cite{EvansPRB2015}, as well as particles with an enhanced magnetic anisotropy due to surface anisotropy, dopants or defects where we choose an arbitrary large value of $k_{\mathrm{c}} = -1 \times 10^{-23}$ J/atom. These two cases provide lower and upper bounds for the attempt frequency in magnetite samples found in nature where imperfections can be a significant source of magnetic anisotropy.


\begin{figure*}[!bt]
     \centering
     \includegraphics[width=1.98\columnwidth,angle=0]{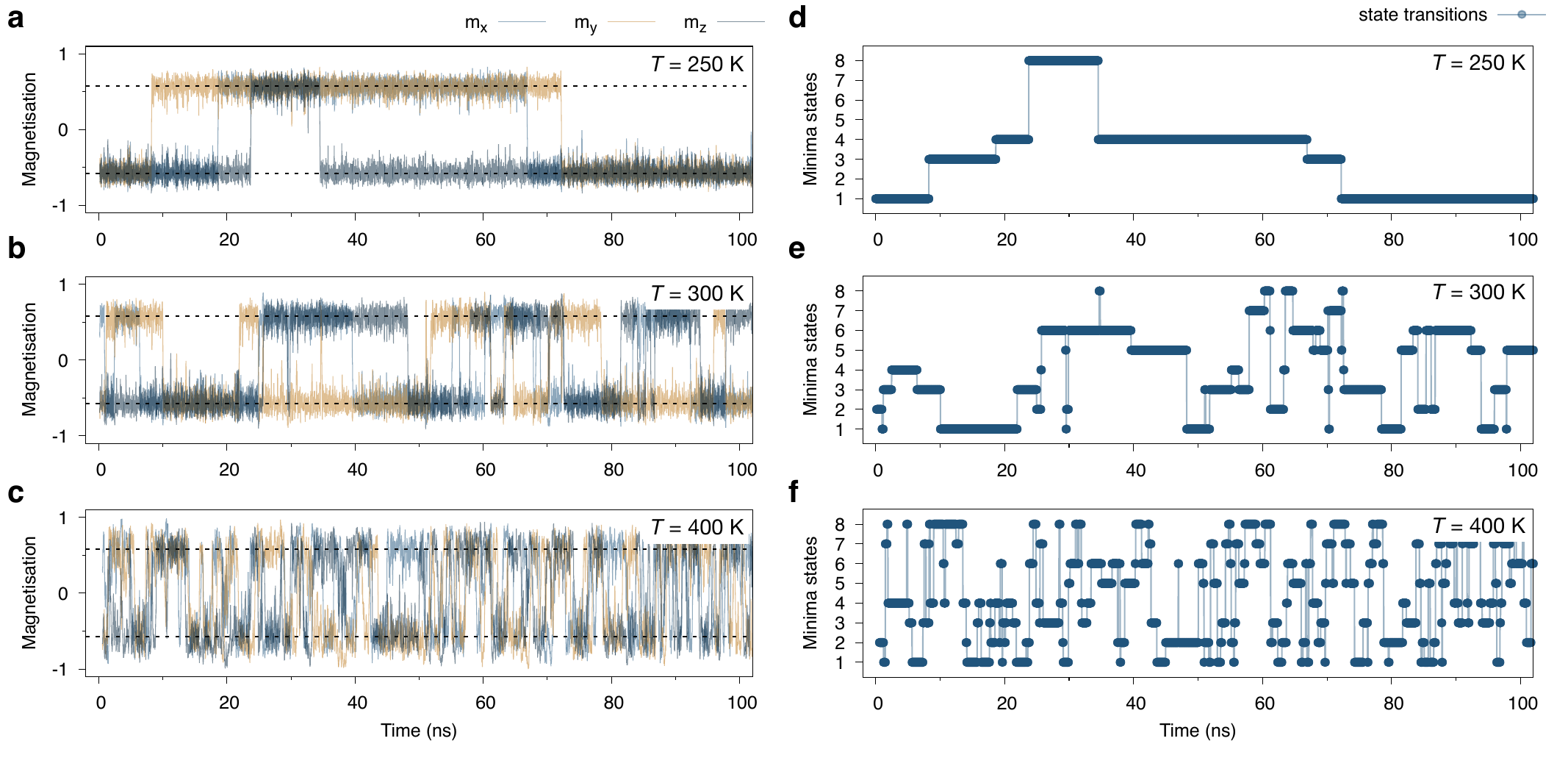}
     \caption{Illustrative Langevin dynamics calculations of the time dependence of the magnetisation for temperatures of (a) $T = 250$ K, (b) $T = 300$ K and (c) $T = 400$ K for a fixed particle size ($D=16$nm). Different coloured lines represent the Cartesian components of the normalized magnetization versus time and horizontal black dashed lines represent the minimum energy state directions for the negative cubic anisotropy energy. The magnetisation exhibits telegraph noise where the magnetisation is metastable in the minimum. The associated magnetic minimum is extracted from the time-dependent magnetisation using the geofencing approach. The corresponding time-evolution of each minimum state (labelled 1-8) for each temperature is shown in panels (d-f). This allows an accurate calculation of the relaxation time $\tau$ for all particle sizes and temperatures.}
     \label{TempVsTime}
\end{figure*}

Here we study truncated octahedral faceted nanoparticles, one the most common morphologies found in natural samples and laboratory manufactured crystals\cite{doi:10.1063/1.4901008,LIU20062979,https://doi.org/10.1029/2021GC009770,D0NR02189J}. A characteristic truncated octahedral nanoparticle is shown schematically in Fig.~\ref{fig:schematic}(a). 
The larger particle sizes considered approach the single domain (SD) regime \cite{nagy_2017,RN17585}, i.e., collective behaviour of the magnetic moments is exchange dominated rather than magnetostatic. It is not computationally possible to simulate magnetite nanoparticle with size above the critical SD diameter ($D_C= 60 \sim 85$ nm)\cite{VXbulter,MORENO2022168495,VXmagnetite}, beyond which vortex domain states dominate. We therefore do not include magnetostatic energy in our calculations. We define the  diameter of the nanoparticles as the diameter of the initial spherical nanoparticle $D$, which is then faceted to obtain octahedral particles. Therefore $D > D_\mathrm{faceted}$, where $D_\mathrm{faceted}$ is the diameter of a sphere with the same volume as the faceted nanoparticle. In this work we stick to the $D$ definition just for particle nomenclature and study particle diameters in the range $D = 8$ nm to $D = 28$ nm, and temperature ranges from  $150$ K to  $400$ K. Each particle size and temperature have been simulated for a maximum of $1000$ ns. The time evolution of the magnetic moments is given by the stochastic Landau-Lifshitz-Gilbert (sLLG) equation applied at the atomic scale \cite{Ellis} with a damping constant of $\lambda = 0.063$ at all spin sites extracted from ferromagnetic resonance measurements~\cite{4f9d0f6c8b614c83b6f2d46a108f9945}. The calculations have been carried out using the \vampire software package\cite{vampireURL,vampire}.


The dynamics of magnetic nanoparticles are highly dependent on the particle size and temperature. Characteristic dynamics of a $D = 16$ nm nanoparticle are shown in Fig.~\ref{TempVsTime}(a)-(c) for different simulation temperatures.
Although calculated to 1000 ns, for clarity of the dynamics we only display the first $100$ ns. The system exhibits telegraph noise as the magnetisation is confined to one of the eight energy minima for a certain period of time. For all temperatures the fluctuations are rapid indicating superparamagnetic behaviour on an experimental timescale. At $T=250K$  the magnetic anisotropy energy dominates the system such that the rate of magnetisation switching from one equilibrium state to another occurs infrequently within the 100ns time scale. However, increasing the temperature by $50$K to RT and then by a further $100$K, the rate of magnetisation switching increases by more than order of magnitude for the same particle size. Increasing the particle volume for a given temperature produces the opposite effect as temperature does.



To accurately determine the attempt frequency it is first necessary to obtain a definitive value of the mean relaxation time, $\tau$. 
Here we use an alternative approach using geofencing, where the magnetization is associated with one of the eight local minima and is associated with that minimum until transitioning to a different minimum. Here we use a criterion that the magnetisation $\mathbf{m}$ is associated with a new minimum state $\alpha$ when $\mathbf{m} \cdot \mathbf{e}_\alpha > 0.91$, where $\mathbf{e}_\alpha$ is the easy direction of the local minimum. This allows for large fluctuations around the minimum state without defining a transition. This approach allows an accurate estimation of the number of transitions $N$ over the simulation, and we express the characteristic relaxation time as $\tau = t/N$ where $t= 1$ \textmu s is the total simulation time.
The associated magnetic states extracted from the dynamics in Fig.~\ref{TempVsTime}(a)-(c) using the geofencing approach are shown in Fig.~\ref{TempVsTime}(d)-(f).
Collating all of the data for different nanoparticle sizes and temperatures we are able to determine the size and temperature dependence of the relaxation rate and extract the attempt frequency $f_0$. 

\begin{figure}[!tb]
    \centering\includegraphics[width=1.0\columnwidth,angle=0]{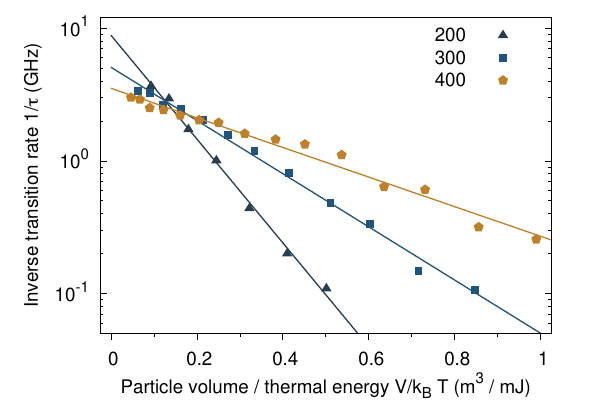}
    \caption{Plot of mean inverse transition rate obtained for each particle size for temperatures of 200K, 300K and 400 K for the case of high anisotropy particles. Solid lines are linear fits for these results. The intercept corresponds to the attempt frequency and the slope  to the energy barrier. The clear linearity of the data indicates a strong correspondence with the Arrhenius-N\'eel law. As the temperature increases the transition rate of each particle increases shown by a decrease in the gradient, while the attempt frequency also decreases with increasing temperature, as shown by the intercept.}
    \label{fig:Arrheniouslaw}
\end{figure}

\begin{figure}[!tb]
    \centering
    \includegraphics[width=1.0\columnwidth,angle=0]{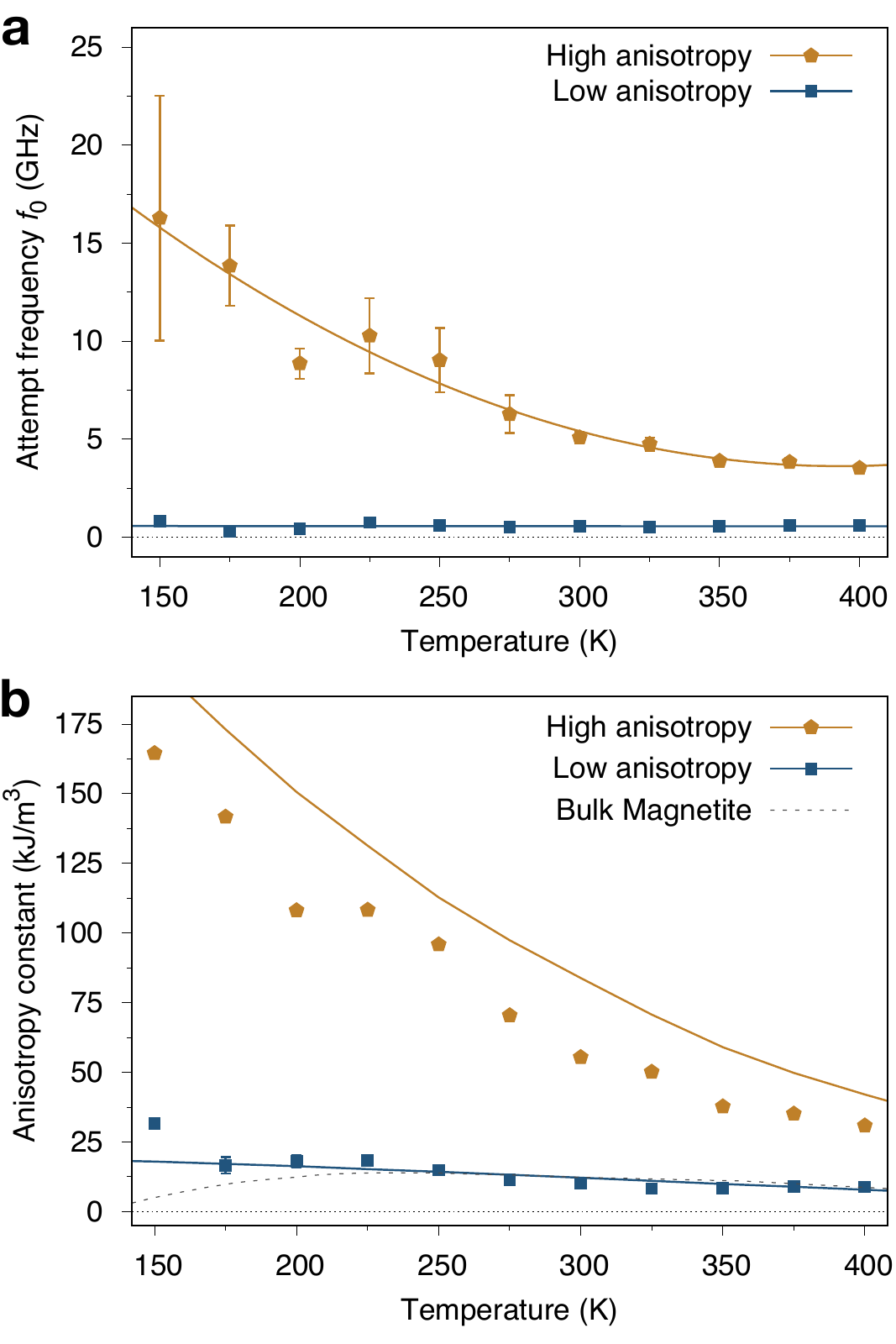}
    \caption{Temperature dependence of the calculated attempt frequency (a) and magnetic anisotropy constant (b) extracted from fits to Eq.~\ref{eq:fit}. Solid lines serve as a guide to the eye for the attempt frequency, and fits to the Zener-Akulov-Callen-Callen (ZACC) scaling of the magnetic anisotropy calculated from $|m|(T)$ for a (20 nm)$^3$ nanoparticle with periodic boundary conditions and assuming a scaling of $K_\mathrm{c}(|m|) = m^{10}$. The error bars are extracted from the fit to the data. The temperature-dependent anisotropy for bulk magnetite is shown by the black dashed line. 
    }\label{Results}
\end{figure}

The volume dependence of $f$ is displayed in Fig~\ref{fig:Arrheniouslaw} for the high anisotropy case and three representative temperatures with a fit of the form
\begin{equation}
    \ln{\left(\frac{1}{\tau}\right)} = \ln{f} = \ln{f_0} - E_\mathrm{B}\frac{V}{\kB T}
    \label{eq:fit}
\end{equation}
where the intercept is the attempt frequency and the gradient gives the energy barrier. Additional data for low anisotropy and a wider range of temperatures is available in supplementary Figures S2-S3. For each temperature $f$ exhibits a linear behaviour with volume that has important implications. Firstly, the linear nature of the data shows that the influence of the particle volume on the attempt frequency value $f_0$ is negligible and its temperature dependence $f_0(T)$ can be determined from the intercept of a linear fit for each temperature case. Secondly, the temperature dependence of the energy barrier $E_\mathrm{B}$ and magnetic anisotropy energy $K_\mathrm{C}(T) = 12 E_\mathrm{B}(T)$ can be accurately determined from the slope of the linear fit. The extracted values for the temperature dependence of the attempt frequency $f_0 (T)$ and the effective anisotropy energy are presented in Fig.~\ref{Results}(a) and ~\ref{Results}(b) respectively. 

The temperature dependence of the attempt frequency exhibits two different behaviours depending on the anisotropy value. For the low anisotropy case closely reproducing the anisotropy of magnetite at room temperature, the attempt frequency is temperature independent with a value of $f_0(300K) = 0.562 \pm 0.059$ GHz, i.e. half of the widely accepted value for magnetite $f_0$. In contrast, for the high anisotropy case $f_0$ decreases with increasing $T$ but with an asymptotic like behaviour at high temperatures. The variation in $f_0$ is however small, less than an order of magnitude in a $250$K temperature window. In this case $f_0 (300K) \sim 5$ GHz. The temperature dependence of the extracted anisotropy energy shown in Fig.~\ref{Results}(b) almost exactly follows the $K(m) = K_0m^{10}$ Zener-Akulov-Callen-Callen scaling law for cubic anisotropy in both high and low anisotropy cases~\cite{ZenerPR1954,AkulovZP1936,Callen1966TheLaw}. $K_0$ is the anisotropy value at zero Kelvin. For the high anisotropy case there is a systematic reduction in the calculated anisotropy energy compared to the exact scaling for a large system (20 nm)$^3$ used to calculate $m(T)$, likely due to finite size effects and the smaller particle size. 


In this work we have simulated the dynamics of single-domain magnetite nanoparticles with cubic anisotropy in a range of sizes and temperatures using atomistic spin dynamics. The temperature range spans below and above room temperature, covering the most relevant cases for paleaomagnetic and magnetic hyperthermia studies. We have studied the cases for low and high anisotropy representing fully passivated Fe$_3$O$_4$ nanoparticles and particles with an enhanced anisotropy arising from either surface anisotropy or dopants such as Co. We find that the attempt frequency shows a different characteristic temperature-dependence depending on the magnetic anisotropy energy of the particles, highlighting the importance of the absolute value of the anisotropy in the dynamic behaviour of magnetic nanoparticles, affecting both the attempt frequency and the energy barrier, in disagreement with previous analytical theories~\cite{Brown,neel1949}. We find a room-temperature attempt frequency for bulk magnetite of $f_0 = 0.562 \pm 0.059$ GHz which falls in the range of the accepted literature value but is established within a microscopic framework and makes no assumptions on whether barriers are low or high, or on unknown factors such as the Gilbert damping constant. More generally our approach enables a direct microscopic calculation of relaxation times, energy barriers and attempt frequencies of complex magnetic systems with high specificity, for example polygranular thin films, non-collinear antiferromagnets~\cite{Jenkins2019}, magnetic tunnel junctions~\cite{Hayakawa2021}, neodymium permanent magnets with higher order anisotropies, magnetic recording media and thin film heterostructures, providing an additional means to understand and control magnetic relaxation at the nanoscale, with potential applications in probabilistic computing~\cite{Misra2022}.

\section*{acknowledgments}
The authors would like to thank Daniel Meilak and Roy Chantrell for helpful discussions. R.M acknowledges the postdoctoral fellowship program of Conicet Argentina. W.W. would like to acknowledge support from the Natural Environmental Research Council through Grants NE/V001233/1 and NE/ S011978/1. This work was supported by the Engineering and Physical Sciences Research Council (grant number EP/P022006/1) using the ARCHER2 UK National Supercomputing Service (https://www.archer2.ac.uk) and the York Viking cluster.

\bibliography{library}
    \label{fig:S2}

    \centering
    \label{fig:S3}



\end{document}